\documentclass[a4paper,11pt]{amsart}

\usepackage{amssymb,amsmath}
\usepackage{amsfonts}
\usepackage[english]{babel}
\usepackage[latin1]{inputenc}
\usepackage[dvips]{graphicx}
\usepackage{cite}
\newtheorem{prop}{Proposition}
\newtheorem{thm}{Theorem}
\newtheorem{cor}{Corollary}
\providecommand{\prt}[1]{\left( #1 \right)}
\providecommand{\abs[1]}{\left|#1\right|}
\providecommand{\quotes[1]}{``#1"}

\title{Order preservation in a generalized version of Krause's opinion dynamics model}
\author{Julien M. Hendrickx.}

\thanks{J.\ M.\ Hendrickx is with Department of Mathematical
Engineering, Universit\'e catholique de Louvain, Avenue Georges
Lemaitre 4, B-1348 Louvain-la-Neuve, Belgium; {\tt\small
julien.hendrickx@uclouvain.be}}

\begin{document}

\maketitle

\begin{abstract}
Krause's model of opinion dynamics has recently been the object of
several studies, partly because it is one of the simplest
multi-agent systems involving position-dependent changing
topologies. In this model, agents have an opinion represented by a
real number and they update it by averaging those agent opinions
distant from their opinion by less than a certain interaction
radius. Some results obtained on this model rely on the fact that
the opinion orders remain unchanged under iteration, a property
that is consistent with the intuition in models with simultaneous
updating on a fully connected    communication topology.\\
Several variations of this model have been proposed. We show that
some natural variations are not order preserving and therefore
cause potential problems with the theoretical analysis and the
consistence with the intuition. We consider a generic version of
Krause's model parameterized by an \quotes{influence function}
that encapsulates most of the variations proposed in the
literature. We then derive a necessary and sufficient condition on
this function for the opinion order to be preserved.
\end{abstract}

\section{Introduction}

\begin{figure}
\centering
\includegraphics[scale = .5]{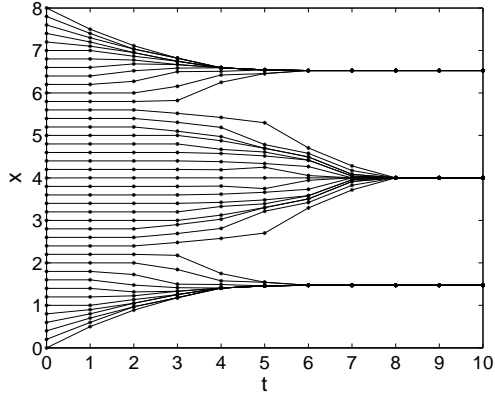}

\caption{Evolution of 41 opinions initially equidistantly
distributed on $[0,8]$ during 10 iterations. At each time $t$,
each opinion is obtained from those at time $t-1$ by the model
given in (\ref{eq:def_Krause_normal}), with
$r=1$.}\label{fig:ex_normal}
\end{figure}

Dynamics of opinions and propagation of beliefs are the object of
many studies in the literature. Agents have an opinion which can
be a continuous value \cite{DeGroot:1974, Fortunato:2005,
Ben-naimKrapivskyRedner:2003,
Ben-naimKrapivskyVasquezRedner:2003,Lorenz:2007} or restricted to
discrete or even binary sets \cite{Galam:1999,Sznajd:2000}. The
evolution of the agents' opinions is influenced by the opinions of
other agents, which typically are their neighbors on some fixed
graph
\cite{DeGroot:1974,Sznajd:2000,BattistonBonabeauWeisbuch:2003} or
are randomly selected at each iteration
\cite{KrapivskyRedner:2003,DeffuantNeauAmblardWeisbuch:2000}. The
originality of Krause's model proposed in 1997 \cite{Krause:1997},
and also known has Hegselmann-Krause model after
\cite{HegselmannKrause:2002}, is that the interaction graph is not
fixed or randomly defined, but depends on the agents opinions in a
deterministic way: two agents influence each other if their
opinion are not too different. Formally, agents have a value
$x_i\in \Re$ interpreted as their opinion on some subject, and
they update it synchronously at every time-step by taking a new
opinion $x_i'$ defined by
\begin{equation}\label{eq:def_Krause_normal}
x_i' = \frac{\sum_{j:\abs{x_i-x_j}\leq r}
x_j}{\abs{\{j:\abs{x_i-x_j}\leq r\}}},
\end{equation}
where the vision range $r$ is a pre-specified constant, and
$\abs{\{j:\abs{x_i-x_j}\leq r\}}$ is the number of agents whose
opinions differ from $x_i$ by at most $r$. Note that this model
presents similarities with the non-deterministic model of Deffuant
et al. \cite{DeffuantNeauAmblardWeisbuch:2000}. Krause's model has
recently been the subject of a wide study
\cite{Krause:2000,HegselmannKrause:2002,Lorenz:2006,
BlondelHendrickxTsitsiklis:2007_ECC,
BlondelHendrickxTsitsiklis:2007_journal,
Fortunato:2004,Fortunato:2005_treshold,
FortunatoLatoraPluchinoRapisarda:2005,Hendrickx:2008phdthesis} due
inter-alia to the fact that it is one of the simplest multi-agent
model involving position-dependent topologies, much simpler for
example than the famous Vicsek swarming model
\cite{VicsekCzirolBenjacobCohenSchchet:1995}. Figure
\ref{fig:ex_normal} shows an example of opinions evolving
according to (\ref{eq:def_Krause_normal}) for 10 iterations. The
opinions converge in finite time to opinion clusters separated by
more than $r$ as shown in \cite{Dittmer:2001} and also in
\cite{Lorenz:2005} in a more general context. The exact distance
between clusters at equilibrium has actually a more complex
behavior, which is studied for
example in \cite{BlondelHendrickxTsitsiklis:2007_ECC,
BlondelHendrickxTsitsiklis:2007_journal, Hendrickx:2008phdthesis}.
One can also see that the opinion order is preserved at each
iteration, if $x_i\leq x_j$ then $x_i'\leq x_j'$. This property
proved in \cite{Krause:2000} is consistent with the intuition that
agent opinions that evolve in a one-dimensional space according to
the same rules and whose communications are not artificially
restrained to an arbitrary topology should not cross each other.
Moreover, the proofs of several important properties of Krause's
model explicitly use the fact that the opinion order is preserved
\cite{BlondelHendrickxTsitsiklis:2007_ECC,
BlondelHendrickxTsitsiklis:2007_journal, Dittmer:2001,
Hendrickx:2008phdthesis}.\\

\begin{figure}
\centering
\begin{tabular}{c}
\begin{tabular}{cc}
\includegraphics[scale = .4]{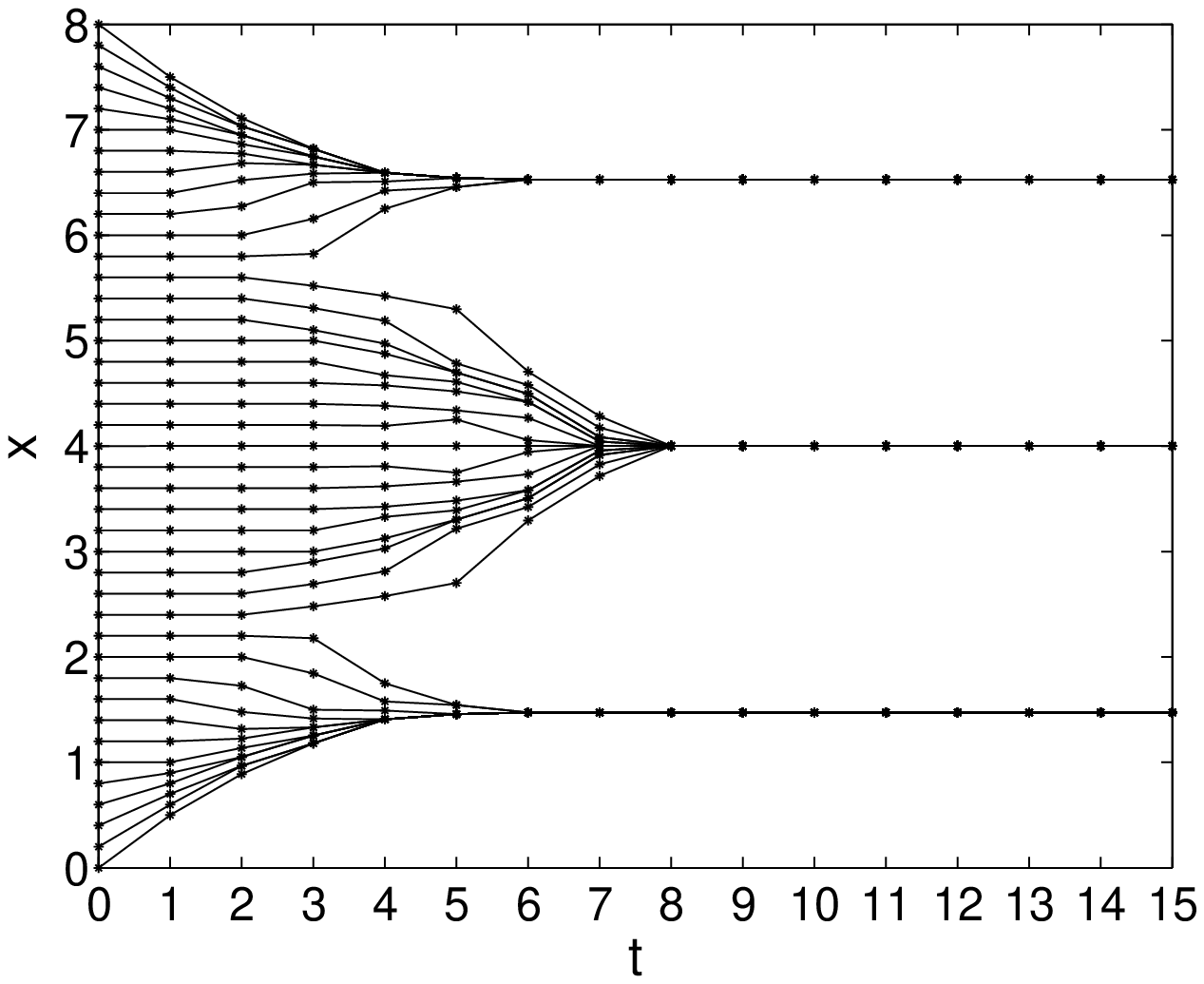}&\includegraphics[scale = .3]{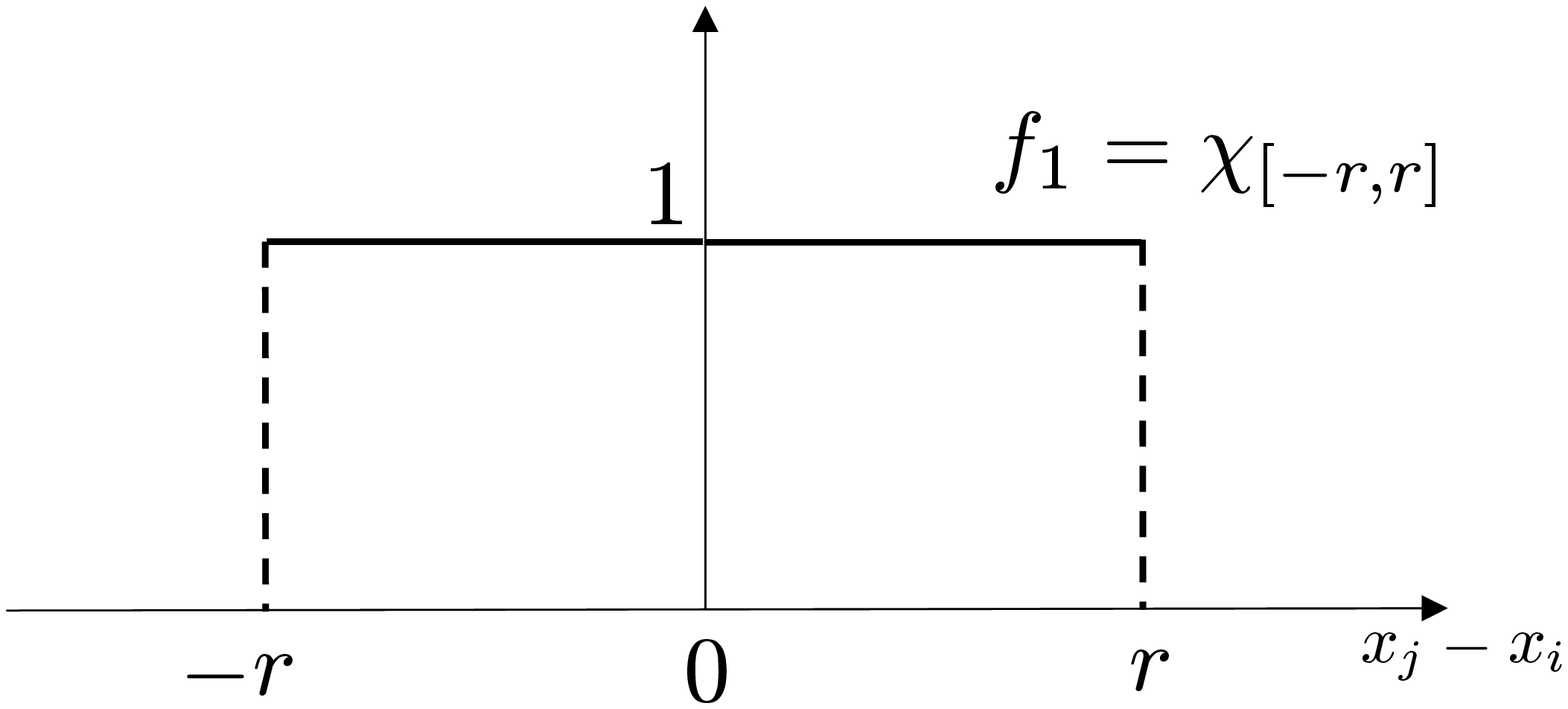}
\end{tabular}\\
(a)\\
\begin{tabular}{cc}
\includegraphics[scale = .4]{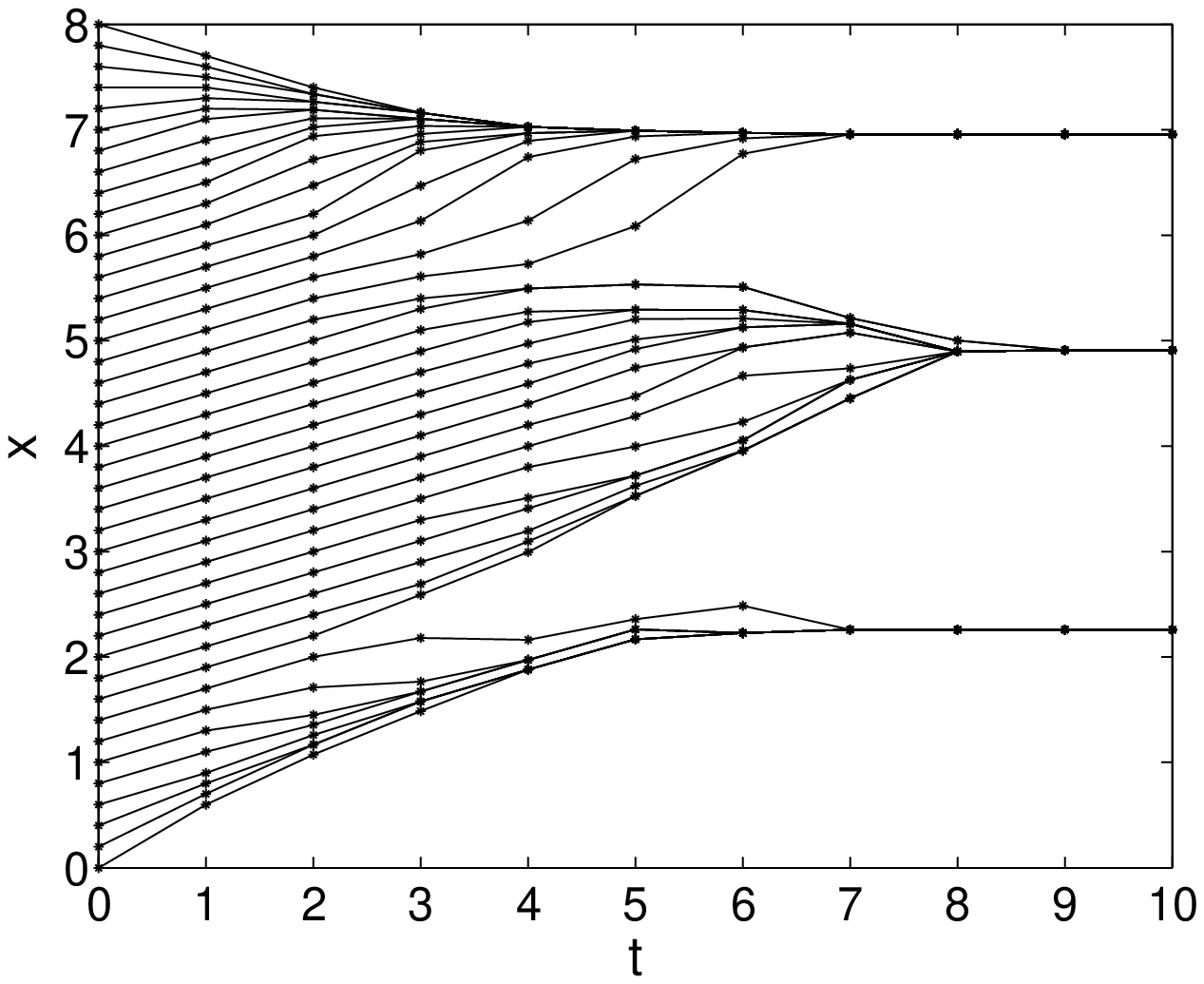}&\includegraphics[scale = .3]{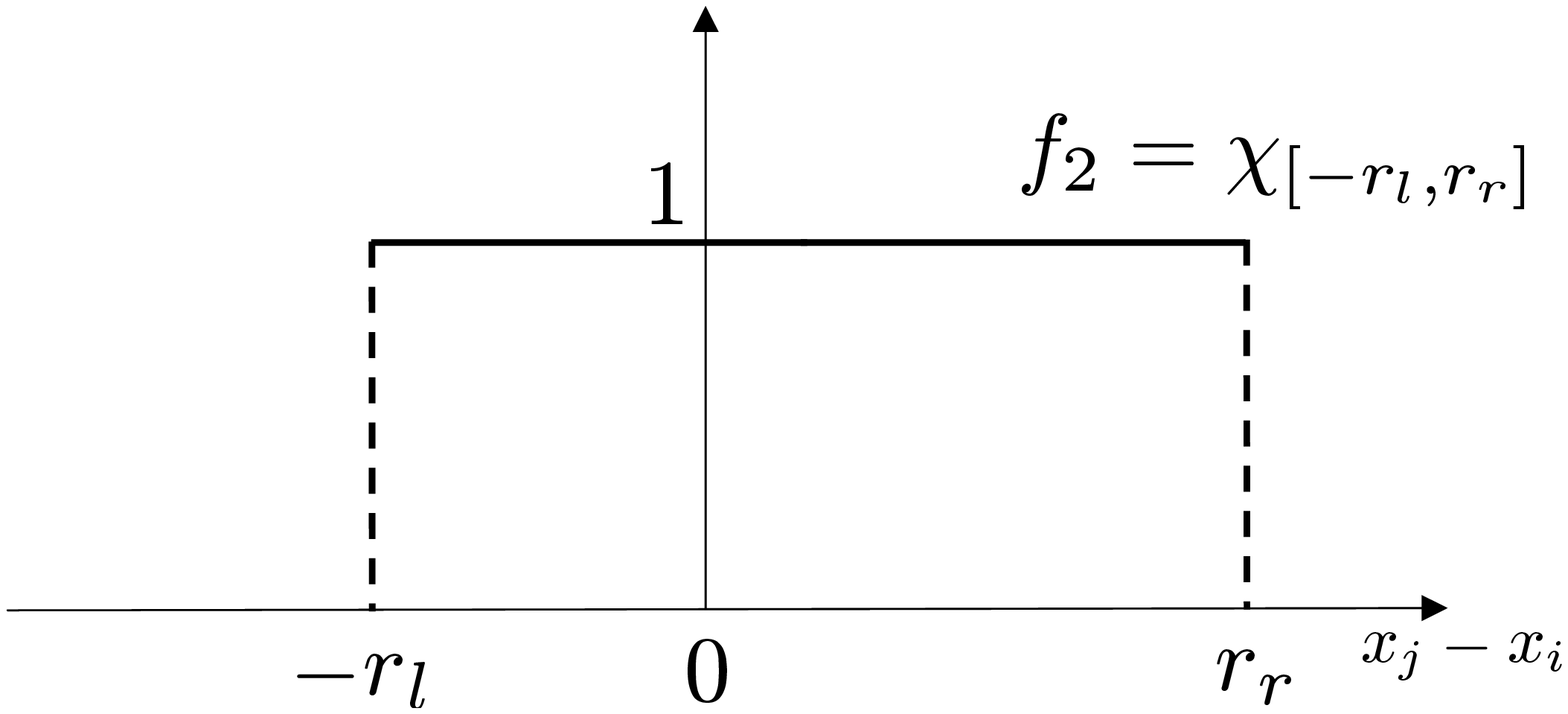}
\end{tabular}\\
(b)\\
\begin{tabular}{cc}
\includegraphics[scale = .4]{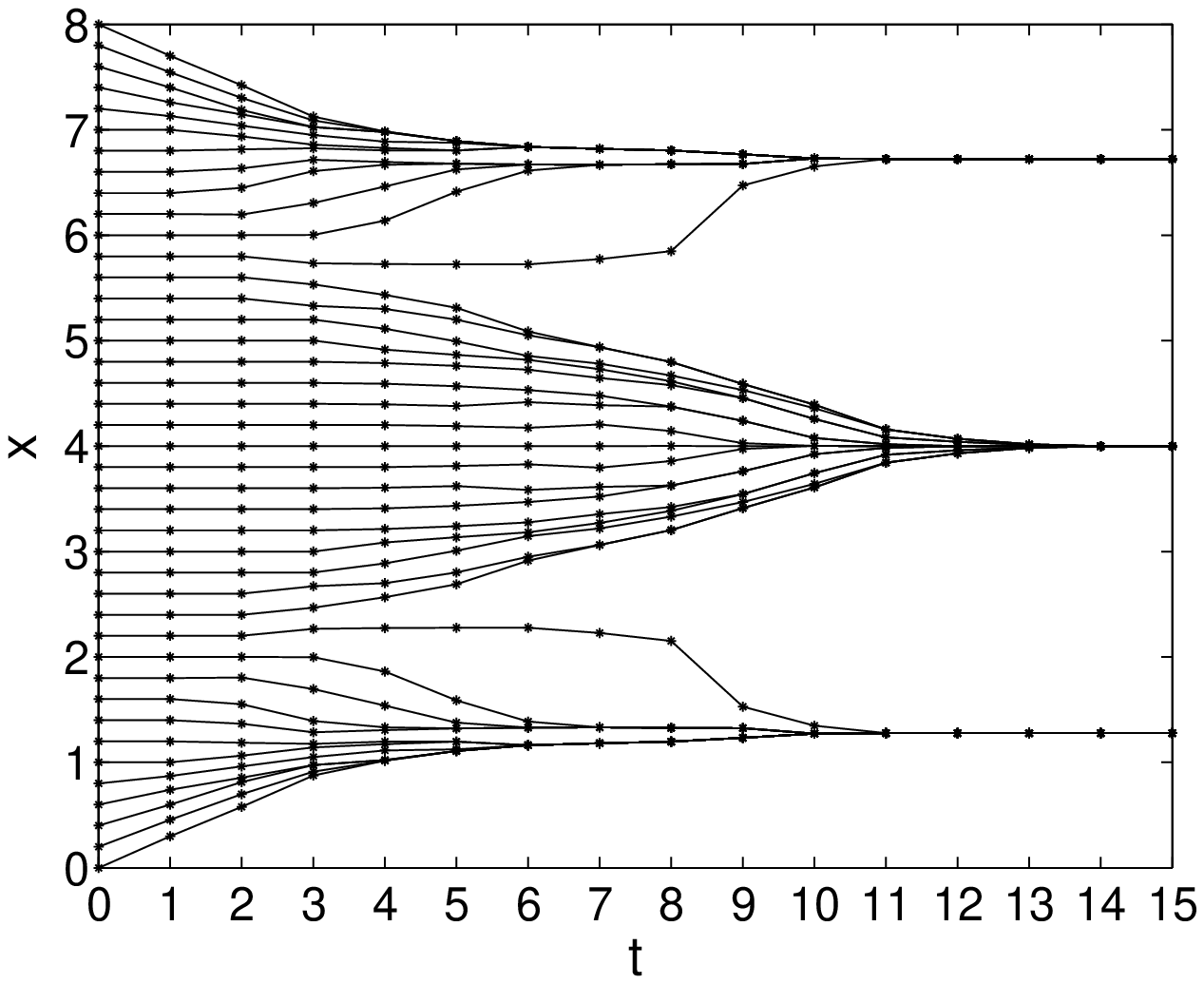}&\includegraphics[scale = .3]{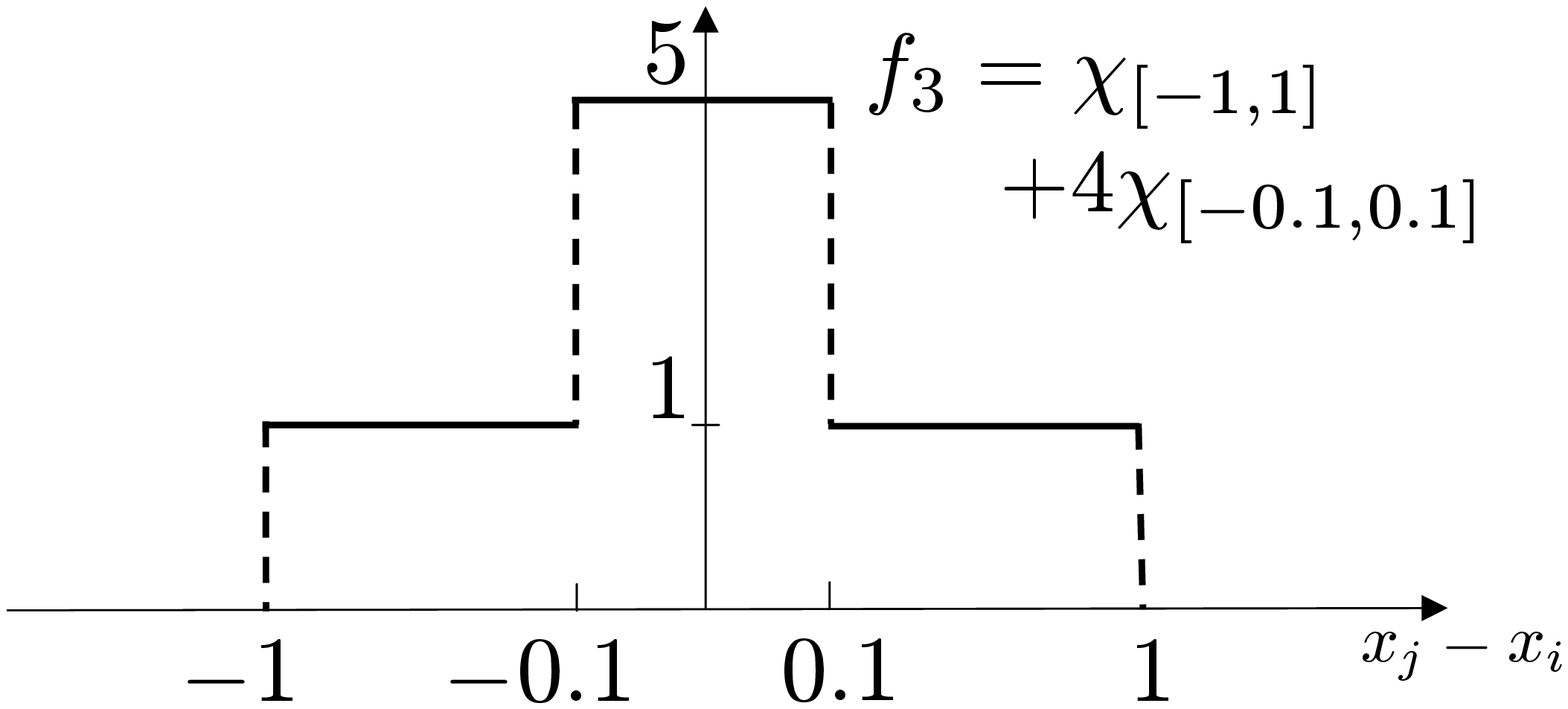}
\end{tabular}\\
(c)\\
\end{tabular}
\caption{Evolution with time of 41 opinions initially
equidistantly distributed on $[0,8]$. The opinions follow the
model (\ref{eq:def_Krause_normal}) with $r=1$ in (a), the model
(\ref{eq:def_Krause_asym}) with $r_l=\frac{5}{4},r_r=\frac{3}{4}$
in (b) and the model (\ref{eq:def_Krause_centerweight}) in (c).
The influence functions describing the respective models are also
represented.}\label{fig:various_op_function}
\end{figure}

\begin{figure}
\centering
 \includegraphics[scale = .5]{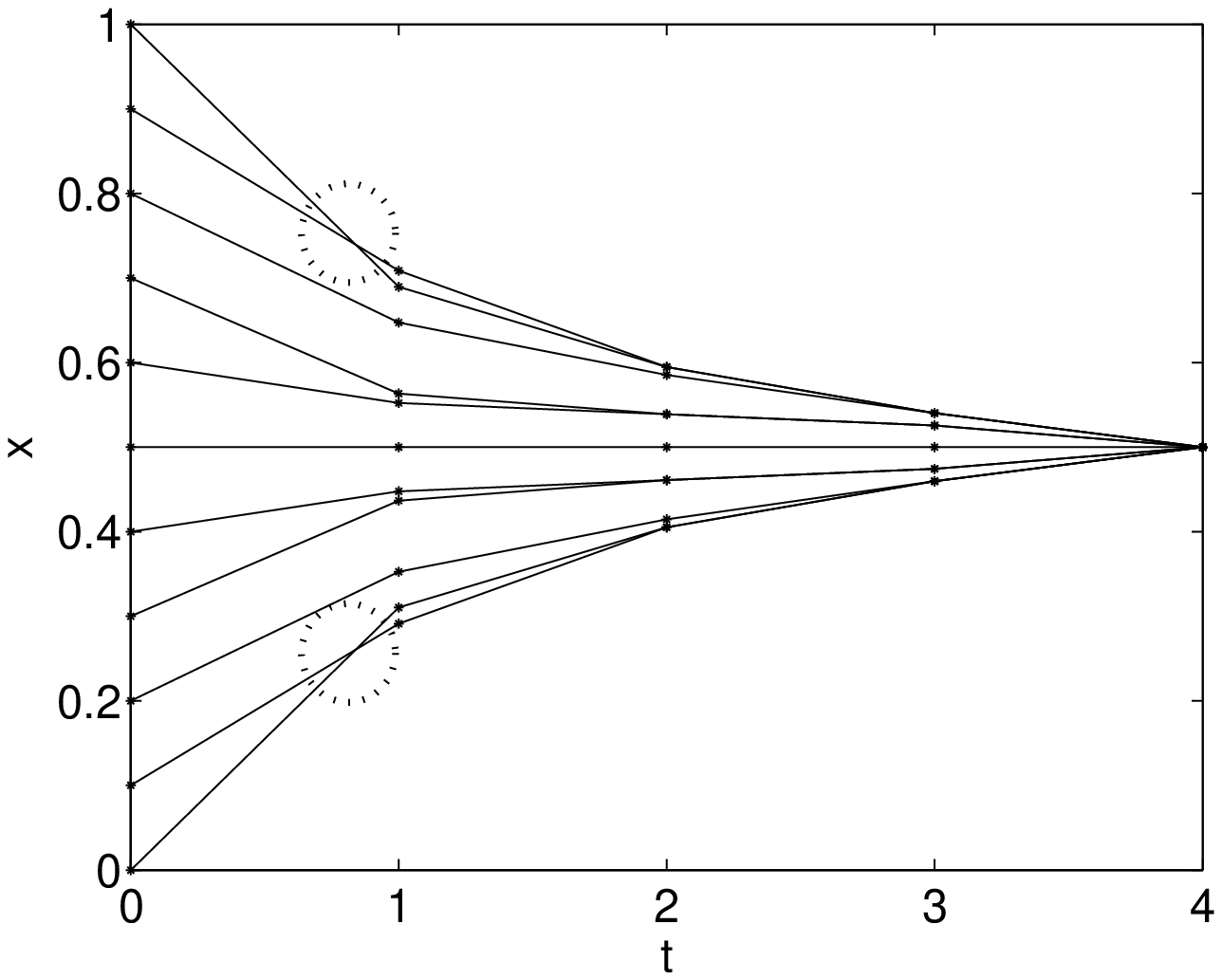}
\caption{Evolution with time of 11 opinions initially
equidistantly distributed on $[0,1]$ and following the model
(\ref{eq:def_Krause_centerweight}). The order of opinions is not
preserved between $t=0$ and $t=1$.}\label{fig:ex_no_order}
\end{figure}

Several extensions of Krause's model have been introduced in the
literature. Asymmetric behavior are for example considered in
\cite{HegselmannKrause:2002}. An agent $i$ takes $j$ into account
if $x_j \in [x_i - r_l,x_i+r_r]$, and the update rule is then
\begin{equation}\label{eq:def_Krause_asym}
x_i' = \frac{\sum_{j: -r_l\leq x_j-x_i\leq r_r} x_j}{\abs{\{j:
-r_l\leq x_j-x_i\leq r_r\}}}.
\end{equation}
Figure \ref{fig:various_op_function}(b) shows the evolution of
opinions according to this model for $r_l = \frac{3}{4}$ and $r_r
= \frac{5}{4}$. Unsurprisingly, opinions converge to clusters with
higher value than in the symmetric case $r_l=r_r=r=1$ represented
in Figure \ref{fig:various_op_function}(a) for comparison. In
another extension of the model proposed in
\cite{Lorenz:2007phdthesis}, the author proposes to weight the
agent opinions depending on the distance separating them. One can
for example weight by 1 the opinions that are at distances between
$\frac{1}{10}$ and $1$ and by 5 those at distance at most
$\frac{1}{10}$:\\
\begin{equation}\label{eq:def_Krause_centerweight}
x_i' = \frac{\sum_{j:\frac{1}{10}<\abs{x_i-x_j}\leq 1} x_j+
\sum_{j:\abs{x_i-x_j}\leq \frac{1}{10}} 5
x_j}{\abs{\{j:\frac{1}{10}<\abs{x_i-x_j}\leq
1\}}+5\abs{\{j:\abs{x_i-x_j}\leq \frac{1}{10}}\}}.
\end{equation}
Opinions evolving according to this rule are represented in Figure
\ref{fig:various_op_function}(c), where it can be seen that
convergence is slower than with the usual model
(\ref{eq:def_Krause_normal}). Although this model seems to be a
natural one, Figure \ref{fig:ex_no_order} shows that it does not
necessarily preserve the opinion order. This renders its potential
validity questionable, as the possibility for opinions evolving
according to the same rules to cross each other can be subject to
debate. Also, it makes the analysis of such a model more
challenging. The analysis of the initial model
(\ref{eq:def_Krause_normal}) in
\cite{BlondelHendrickxTsitsiklis:2007_ECC,
BlondelHendrickxTsitsiklis:2007_journal}  uses indeed extensively
the order preservation property, and its adaptation to models such
as (\ref{eq:def_Krause_centerweight}) can thus be uneasy.\\

In order to characterize those variations of Krause's model that
preserve the opinion order, we consider the following generic
model already suggested in \cite{Lorenz:2007phdthesis} and
encapsulating all value-independent variations of Krause's model.
The generic update rule is
\begin{equation}\label{eq:def_gen_krause}
x_i' = \frac{\sum_{j}f(x_j-x_i) x_j}{\sum_{j}f(x_j-x_i)},
\end{equation}
where $f:\Re\to \Re^+$ is a nonnegative function whose
support\footnote{The support of a non-negative function is the set
on which it takes positive values.} is a positive length interval
containing 0. We call the function $f$ an \emph{influence
function}. In particular, the model (\ref{eq:def_Krause_normal})
is obtained by taking $f_1 = \chi_{[-r,r]}$ as influence function,
where the $\chi_S$ defined for any set $S$ is the \emph{indicator
function} of $S$, that takes the value $1$ on $S$ and 0 everywhere
else. The asymmetric model (\ref{eq:def_Krause_asym}) corresponds
to $f_2= \chi_{[-r_l,r_r]}$, and the model
(\ref{eq:def_Krause_centerweight}) to $f_3= \chi_{[-1,1]} + 4
\chi_{[-\frac{1}{10},\frac{1}{10}]}$ as represented in Figure
\ref{fig:various_op_function}. Lorenz also proposes in
\cite{Lorenz:2007phdthesis} a time varying function
$f(y)=e^{-\abs{\frac{y}{r}}^t}$, but we do not consider
time-varying functions here. Note finally that the idea of
describing various extensions of an opinion dynamics model via an
influence function has also been applied to the model introduces
by Deffuant et al. in \cite{DeffuantNeauAmblardWeisbuch:2000} as
mentioned in \cite{WeisbuchDeffuantAmblard:2005}.\\

We say that an influence function is \emph{order preserving} if
for any state vector $x\in \Re^n$, the updated vector obtained by
equation (\ref{eq:def_gen_krause}) satisfies $x'_i\leq x'_j$ for
any $i,j$ such that $x_i\leq x_j$. Note that when a function is
not order preserving, there may very well exist many different
initial conditions such that no crossing takes place along the
evolution of the system. However, there always exists at least one
initial state vector $x\in \Re^n$ such that the order of two
opinions is inverted after one opinion update. If no such vector
exist, then the function is order preserving. The function $f_3$
in Figure \ref{fig:various_op_function} is for example not order
preserving as shown in Figure \ref{fig:ex_no_order}, while it can
be proved that the functions $f_1$ and $f_2$ are order preserving.\\

We give in Section \ref{sec:alg_cond} a simple necessary and
sufficient condition for a function to be order preserving. We
show in Section \ref{sec:log_conc} how this condition relates to
the log-concavity of the influence function. We close the paper in
Section \ref{sec:ccl_openquestions} by the concluding remarks and
the mention of two open questions.

\section{Algebraic condition for order preservation}\label{sec:alg_cond}

We first consider a very simple system with $2n+2$ agents. The
agents 1 and 2 have opinions $x_1=a$ and $x_2 =b$ respectively,
for some $b> a$. Among the remaining $2n$ agents, $n$ have an
opinion $a+b$ and $n$ others an opinion $a+b+c$ for some $c> 0$.
We suppose that 1 and 2 both take all other agents into account,
that is, $f(a),f(b),f(a+c),f(b+c)>0$. If $n$ is sufficiently
large, we can neglect the agents $1$ and $2$ in the computation of
$x_1'$ and $x_2'$, which according to (\ref{eq:def_gen_krause})
are given by
\begin{equation*}
\begin{array}{lll}
x'_1 &\simeq &\frac{nf(b)(a+b) + nf(b+c)(a+b+c)}{nf(b)+ nf(b+c)}=
a+b + \frac{c}{1+f(b)/f(b+c)},\\
\\
x'_2 &\simeq &\frac{nf(a)(a+b) + nf(a+c)(a+b+c)}{nf(a)+ nf(a+c)} =
a+b + \frac{c}{1+f(a)/f(a+c)}.
\end{array}
\end{equation*}
So if $\frac{f(a)}{f(a+c)}>\frac{f(b)}{f(b+c)}$, then $x_1'> x_2'$
although $x_1\leq x_2$. As a result, if $f$ is order preserving,
for any $a<b$ and $c>0$, there holds $\frac{f(a+c)}{f(a)}\geq
\frac{f(b+c)}{f(b)}$, for otherwise we could build the
example above.\\

To see that this simple condition is also sufficient for order
preservation, we now consider another system of $n$ agents among
which we select two agents $p$ and $q$ such that $x_q\geq x_p$ (we
may possibly chose $p=q$). We suppose that $f(x_i-x_p)>0$ and
$f(x_i-x_q)>0$ for all agents $i$. Since the system
(\ref{eq:def_gen_krause}) is translation-invariant we assume that
all $x_i$ are nonnegative, and we relabel the agents in such a way
that $x_1 \leq x_2\leq \dots \leq x_n$. The updated values of
$x_p$ and $x_q$ are
\begin{equation*}
x_p'= \frac{\sum_{i=1}^n f(x_i-x_p) x_i} {\sum_{i =1}^n
f(x_i-x_p)},\phantom{aaa}\text{and}\phantom{aaa} x_q'=
\frac{\sum_{i=1} ^nf(x_i-x_q)x_i} {\sum_{i=1}^n f(x_i-x_q)}.
\end{equation*}
As a consequence, $x_q'\geq x_p'$ holds if
\begin{equation*}
\prt{\sum_{i= 1}^n f(x_i-x_q) x_i}\prt{\sum_{i=1}^n f(x_i-x_p)}
\geq \prt{\sum_{i= 1}^n f(x_i-x_p) x_i}\prt{\sum_{i=1}^n
f(x_i-x_q)}
\end{equation*}
holds. This can be rewritten as
\begin{equation*}
\begin{array}{llll}
&\prt{\sum_{i= 1}^{n-1} f(x_i-x_q) x_i}\prt{\sum_{i=1}^{n-1}
f(x_i-x_p)} &+& f(x_n-x_q)x_n \prt{\sum_{i=1}^{n-1} f(x_i-x_p)} \\
+& f(x_n-x_p) \prt{\sum_{i= 1}^{n-1} f(x_i-x_q)  x_i}& +&
f(x_n-x_q)f(x_n-x_p)x_n 
\\
\geq &\prt{\sum_{i= 1}^{n-1} f(x_i-x_p) x_i}\prt{\sum_{i=1}^{n-1}
f(x_i-x_q)} &+& f(x_n-x_p)x_n \prt{\sum_{i=1}^{n-1} f(x_i-x_q)} \\
+&f(x_n-x_q) \prt{\sum_{i= 1}^{n-1} f(x_i-x_p) x_i} &+&
f(x_n-x_p)f(x_n-x_q) x_n.
\end{array}
\end{equation*}
For $n=1$ (and thus $a=b=1$), this relation reduces to $f(0)^2x_1
\geq f(0)^2 x_1$ and is trivially satisfied. Suppose now that it
holds for $n-1$, then it also holds for $n$ provided that
\begin{equation}\label{eq:prooforder:cond_recur}
\begin{array}{lll}
&f(x_n-x_q)x_n \prt{\sum_{i=1}^{n-1} f(x_i-x_p)} &+ f(x_n-x_p)
\prt{\sum_{i= 1}^{n-1}f(x_i-x_q) x_i}\\ \geq &f(x_n-x_p)x_n
\prt{\sum_{i=1}^{n-1} f(x_i-x_q)} &+ f(x_n-x_q) \prt{\sum_{i=
1}^{n-1} f(x_i-x_p)x_i}. \end{array}
\end{equation}
holds. Reorganizing the terms of (\ref{eq:prooforder:cond_recur})
and dividing them by $f(x_n-x_p)f(x_n-x_q)x_n>0$ yields
\begin{equation*}
\sum_{i=1}^{n-1}
\prt{\frac{f(x_i-x_p)}{f(x_n-x_p)}-\frac{f(x_i-x_q)}{f(x_n-x_q)}}
\geq \sum_{i= 1}^{n-1}
\frac{x_i}{x_n}\prt{\frac{f(x_i-x_p)}{f(x_n-x_p)}-\frac{f(x_i-x_q)}{f(x_i-x_n)}}.
\end{equation*}
Since all $x_i$ are nonnegative and no greater than $x_n$, it is
sufficient for this relation to hold that
$\frac{f(x_i-x_p)}{f(x_n-x_p)}\geq \frac{f(x_i-x_q)}{f(x_n-x_q)}$
holds for all $i$. Since $x_p \leq x_q$, the latter is always true
if $f$ is such that $\frac{f(a+c)}{f(a)}\geq \frac{f(b+c)}{f(b)}$
holds for any $b\geq a$ and $c\geq 0$ for which
$f(a),f(b),f(a+c),f(b+c)>0$. It suffices indeed to take
$a=x_i-x_q$, $b= x_i -x_p$ and $c=x_n-x_i$.\\

Suppose now that there is some $i$ for which $f(x_i-x_p)>0$ and/or
$f(x_i-x_q)>0$ does not hold. Let $J_p$ be the set of agents $i$
such that $f(x_i-x_p)>0$, $J_q$ the corresponding set for $x_q$
and $I = J_p \cap J_q$. If $I = \varnothing$, then any value of
$J_q$ is larger than all values of $J_p$ as the support of $f$ is
an interval, so that $x_q'\geq x_p'$ trivially holds. If
$I=J_p\cup J_q$, we have seen that the necessary condition for
order preservation is sufficient for $x_q'\geq x_p'$ to hold.
Finally, observe that the presence of agents in $J_q\setminus I$
or in $J_p\setminus I$ only increases $x_q'$ or decreases $x_p'$,
so that this condition is still sufficient for $x_q'\geq x_p'$ to
hold. We have thus proved the following result:

\begin{thm}\label{thm:orderpreserve}
An influence function $f:\Re\to \Re^+$ is order preserving if and
only if
\begin{equation}\label{eq:condition_f}
\frac{f(a+c)}{f(a)}\geq \frac{f(b+c)}{f(b)}
\end{equation}
holds for all $a\leq b$, $c\geq 0$ such that
$f(a),f(b),f(a+c),f(b+c)>0$
\end{thm}
Note that the model of Krause can be extended to continuous
distribution of opinions
\cite{BlondelHendrickxTsitsiklis:2007_ECC,
BlondelHendrickxTsitsiklis:2007_journal,
FortunatoLatoraPluchinoRapisarda:2005, Lorenz:2007,
Hendrickx:2008phdthesis} Theorem \ref{thm:orderpreserve} can also
be proved for such systems, replacing sums by integrals, and
assuming that $f$ is
measurable.\\

To understand the intuitive meaning of this result, consider three
agents $i$, $p$ and $q$, with $x_i < x_p < x_q$. The values
$f(x_p-x_i)$ and $f(x_q-x_i)$ represent the \quotes{importance}
given respectively to $p$ and $q$ by $i$ when computing its new
opinion, that is, the weight given by $i$ to the opinions of $p$
and $q$ respectively. The ratio $\frac{f(x_p-x_i)}{f(x_q-x_i)}$ is
thus large if $i$ discriminates $q$ with respect to $p$, i.e.
gives much more importance to $p$ than to $q$, and small
otherwise. Observe now that an agent having an opinion $x_i-c$
would have a \quotes{discriminating ratio}
$\frac{f(x_p-x_i+c)}{f(x_q-x_i+c)}$. Taking $a = x_p-x_i$ and $b =
x_q- x_i$, one can verify that the condition of Theorem
\ref{thm:orderpreserve} implies that an agent with an opinion $x_i
- c$ should discriminate more $p$ from $q$ than an agent with
opinion $x_i$. Since this is true for any $c$ and $x_i$, the
condition of Theorem \ref{thm:orderpreserve} means that the more
remote the opinion of an agent is, the more it should discriminate
$p$ from $q$ (with $x_p < x_q$). This may seem surprising as one
would expect that an agent having an opinion very different from
$x_p$ and $x_q$ should treat them more equally
than an agent having an opinion close to one of them.\\

Using Theorem \ref{thm:orderpreserve}, one can see that the
function $f_3$ in Figure \ref{fig:various_op_function} is not
order preserving, as observed in Figure \ref{fig:ex_no_order}.
Take indeed $a=0$, $b=0.5$ and $c = 0.2$. We have $a\leq b$,
$c\geq 0$, and there holds
\begin{equation*}
\frac{f_3(a+c)}{f_3(a)} = \frac{f_3(0.2)}{f_3(0)} = \frac{1}{5} <
\frac{1}{1}= \frac{f_3(0.7)}{f_3(0.5)} = \frac{f_3(b+c)}{f_3(b)},
\end{equation*}
so that $f_3$ does not satisfy the condition of Theorem
\ref{thm:orderpreserve}. It will be seen later that no
\quotes{reasonable function} is order preserving if it  is
discontinuous somewhere in the interior of its domain. One can
prove on the other hand that the functions $f_1$ and $f_2$ in
Figure \ref{fig:various_op_function} do satisfy these conditions
and are thus order preserving. Proving that a function satisfy
these conditions or finding $a$, $b$ and $c$ that invalid them may
however not always be trivial. For this reason, we show in the
next section that these conditions can be re-expressed in term of
the concavity of $\log f$.

\section{Log-concavity of influence functions}\label{sec:log_conc}
\begin{figure}
\centering
\begin{tabular}{ccc}
\includegraphics[scale=.3]{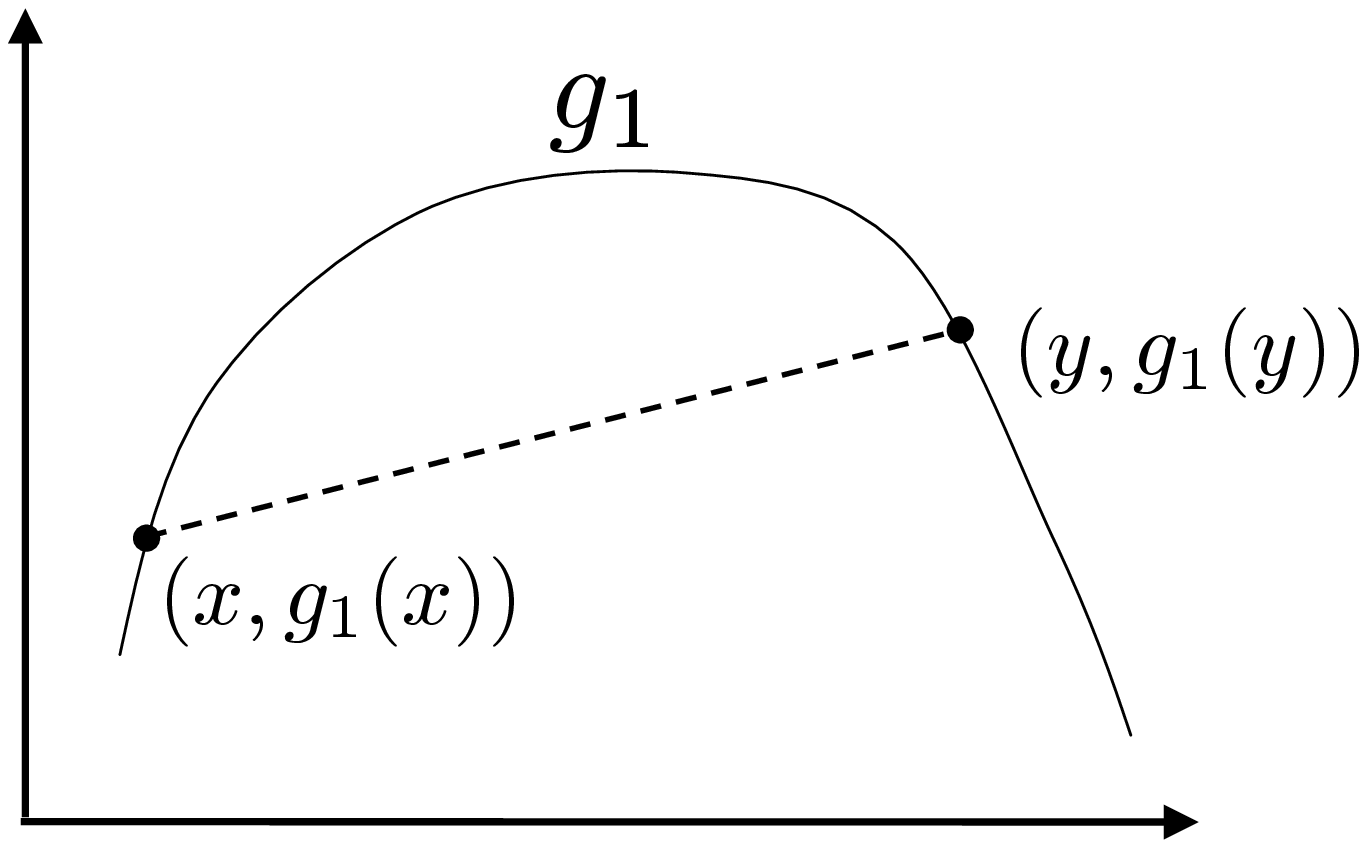}&\includegraphics[scale=.3]{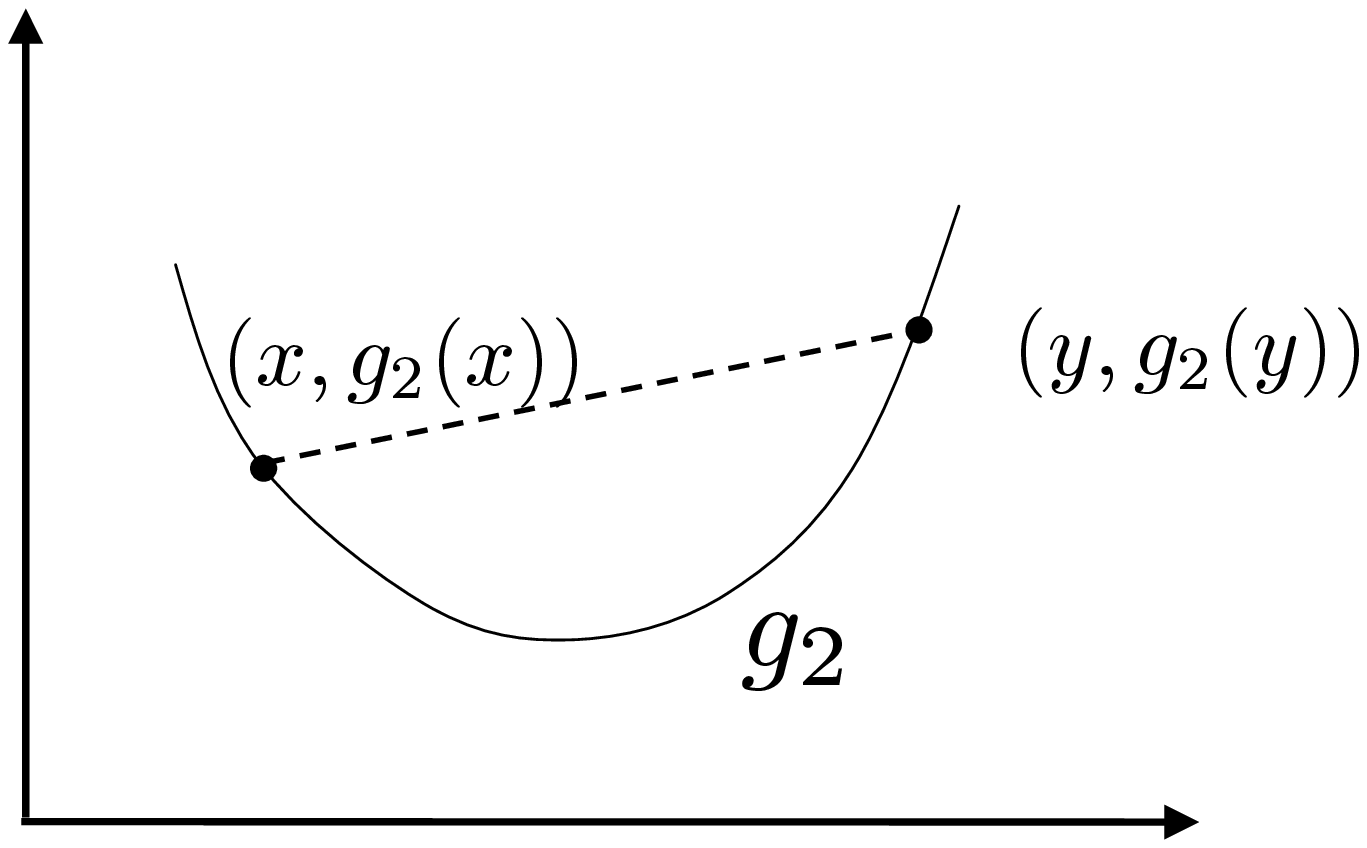}
&\includegraphics[scale=.3]{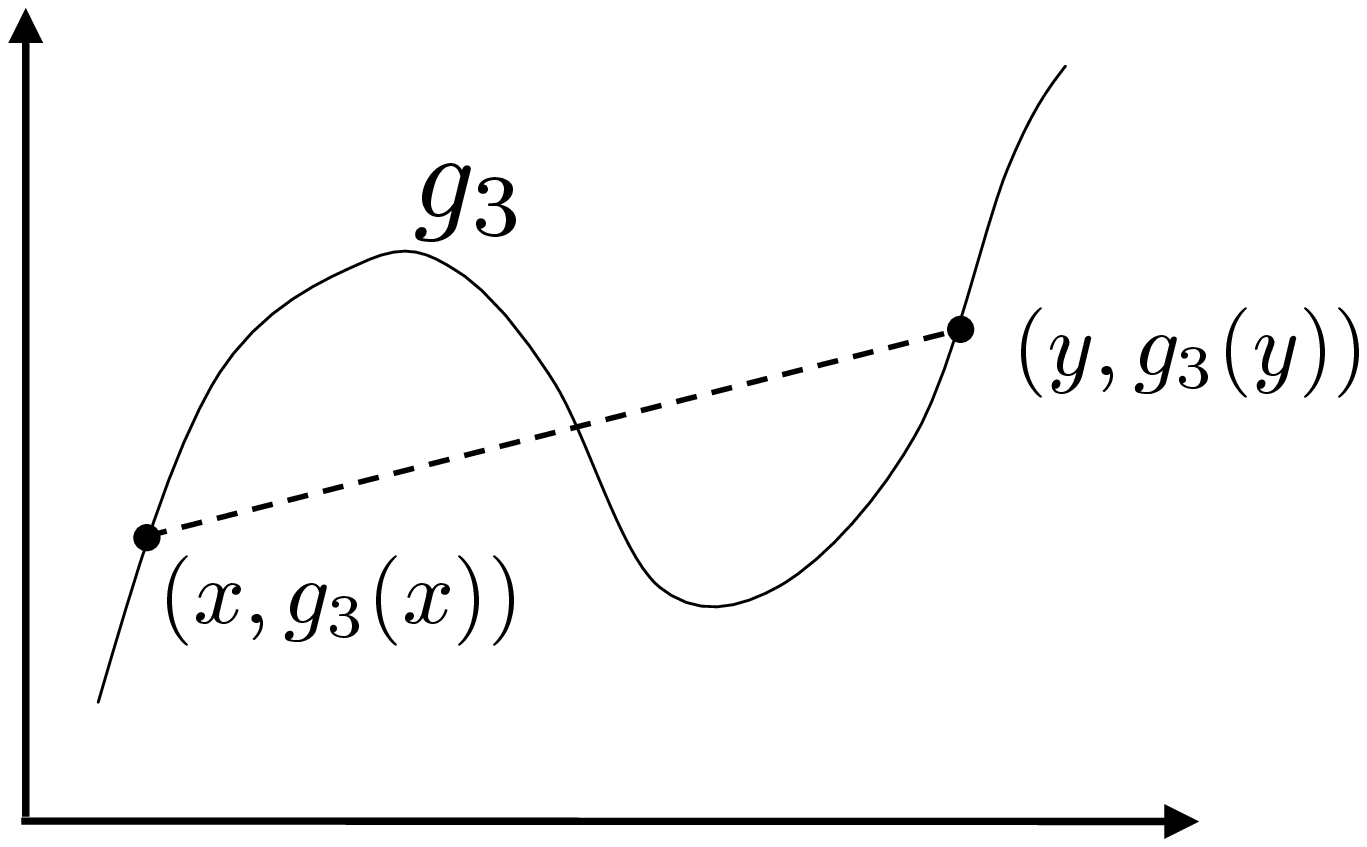}\\ (a) &(b) &(c)
\end{tabular}
\caption{Examples of concave and of non-concave functions. The
function $g_2$ is not concave because for the $x$ and $y$
represented, the line between $(x,g_2(x))$ and $(y,g_2(y))$ is
above the curve of $g_2$. The function $g_3$ is not concave either
because the represented line crosses the curve. On the other hand,
$g_1$ is concave, as for any $x$ and $y$, the line between
$(x,g_1(x))$ and $(y,g_1(y))$ remains below the curve of
$g_1$.}\label{fig:concave_non_concave}
\end{figure}

Remember that a function $g$ is \emph{concave} if for any $x, y$
in its domain, and any $\lambda\in [0,1]$, there holds
\begin{equation*}
g\prt{\lambda x + (1-\lambda) y} \leq \lambda g(x) + (1-\lambda)
g(y).
\end{equation*}
In other words, $g$ is concave if for any $x$ and $y$, the line
between $(x,g(x))$ and $(y,g(y))$ remains below the curve of $g$
and does not cross it. Figure \ref{fig:concave_non_concave} shows
examples of concave and non concave functions.  A concave function
is always continuous on the interior of its domain, that is,
everywhere except possibly on the frontier of its domain of
definition. When a function is differentiable, it is concave if
and only if its derivative is non-increasing. And when it is twice
differentiable, it is concave if and only if its second derivative
is non-positive.\\

We now analyze how the algebraic condition of Theorem
\ref{thm:orderpreserve} can be related to the concavity of $\log
f$, which is often simpler to check. Note that in the sequel we
always implicitly assume that the points at which
$\log f$ is evaluated belong to the support of $f$.\\

Let us first assume that $f$ (and thus also $\log f$) is
differentiable. Taking the logarithm of (\ref{eq:condition_f}), we
see that $f$ is order preserving if and only if, for any $c>0$,
$a\leq b$, there holds
\begin{equation}\label{eq:condition_logf}
\log f(a+c) - \log f(a) \geq \log f(b+c) - \log f(b).
\end{equation}
Taking the limit for $c\to 0$, this condition implies that $(\log
f)'(a)\geq (\log f)'(b)$ for all $a\leq b$, and thus that $\log f$
is concave since its derivative is non-increasing. Similarly, if
$\log f$ is concave, there holds $(\log f)'(a)\geq (\log f)'(b)$
for all $a\leq b$, and one can then show by integrating $(\log
f)'$ that the condition (\ref{eq:condition_logf}) holds for any
$c>0$, so that $f$ is order preserving. As a result, a
differentiable function is order preserving if and only if it is
\emph{log-concave}, that is, if and only if its logarithm is
concave.\\

Many natural influence functions are however not differentiable or
continuous everywhere (see for example Figure
\ref{fig:various_op_function}(c)). We consider thus now an
influence function $f$ on which no smoothness assumption is made.
Suppose first $\log f$ is concave. Then for $b>a$ and $c>0$, there
hold
\begin{equation*}
\begin{array}{lll}
\log f(a+c)&\geq \frac{b-a}{b-a+c} \log f(a) &+ \frac{c}{b-a+c}\log f(b+c),\\
\log f(b) &\geq \frac{c}{b-a+c} \log f(a) &+ \frac{b-a}{b-a+c}\log f(b+c).
\end{array}
\end{equation*}
Adding these two inequalities leads to $\log f(a+c) - \log f(a)
\geq \log f(b+c) - \log f(b)$, or equivalently
$\frac{f(a+c)}{f(a)}\geq \frac{f(b+c)}{f(b)}$. So it follows from
Theorem \ref{thm:orderpreserve} that a log-concave influence
function, that is a function whose logarithm is concave, is always
order
preserving.\\

Conversely, let now  $x<y<z$ be arbitrary points of its support
such that $(y-x)/(z-y)$ is rational. There exist thus two integers
$m,n$ such that
\begin{equation*}
\frac{z-y}{n} = \frac{y-x}{m} =: c >0.
\end{equation*}
If $f$ is order preserving, it follows from Theorem
\ref{thm:orderpreserve} that
\begin{equation*}
\log f(a+c) - \log f(a) \geq \log f(b+c) - \log f(b).
\end{equation*}
holds for any $a<b$ in the support of $f$. So we have
\begin{equation*}
\begin{array}{lllll}
\log f(y)-\log f(x) &= &\sum_{j=1}^{m}\prt{\log f \prt{x+jc}-\log
f\prt{x+(j-1)c}} &\geq &m
\prt{\log f\prt{y+c}-\log f\prt{y}},\\
\log f(z)-\log f(y) &= &\sum_{j=1}^{n}\prt{\log f\prt{y+jc}-\log
f\prt{y+(j-1)c}} &\leq &n \prt{\log f\prt{y+c}-\log f\prt{y}},
\end{array}
\end{equation*}
which implies that
\begin{equation}\label{eq:ratio_concave}
\log f(y)\geq \frac{n}{n+m} \log f(x) +\frac{m}{n+m}\log f(z)
=\frac{z-y}{z-x}\log f(x) + \frac{y-x}{z-x}\log f(z).
\end{equation}
If this holds for any $x$, $y$ and $z$, then $\log f$ is concave.
But remember that (\ref{eq:ratio_concave}) only holds for those
$x,y,z$ for which $(y-x)/(z-x)$ is rational. Nevertheless, if $f$
is continuous on its support, its continuity together with
(\ref{eq:ratio_concave}) implies that it is concave. And, the
following proposition proved in Appendix \ref{sec:appen_proof}
shows that every \quotes{reasonable} order preserving influence
function is continuous on its support:

\begin{prop}\label{prop:if_not_concave}
Let $f$ be an order preserving influence function that is
discontinuous at one point of its support's interior. Then $f$
admits a positive lower bound on no positive length interval, and
is as a consequence discontinuous everywhere on its support.
\end{prop}

The results of this section are summarized in the following
theorem.

\begin{thm}\label{thm:logconcave}
Let $f:\Re \to \Re^+$ be an influence function. If $\log f$ is
concave, then $f$ is order preserving. And if $f$ is order
preserving and admits a positive lower bound on at least one
positive-length interval or is continuous at one point of its
support, then $\log f$ is concave.
\end{thm}

Using the fact that the logarithm of a concave function is also
concave, we obtain the following Corollary.

\begin{cor}\label{cor:concave}
Let $f:\Re \to \Re^+$ be an influence function. If the restriction
of $f$ to its support (i.e. the set on which it takes positive
values) is concave, then it is order preserving.
\end{cor}

\begin{figure}
\centering \begin{tabular}{cc}
\includegraphics[scale=.25]{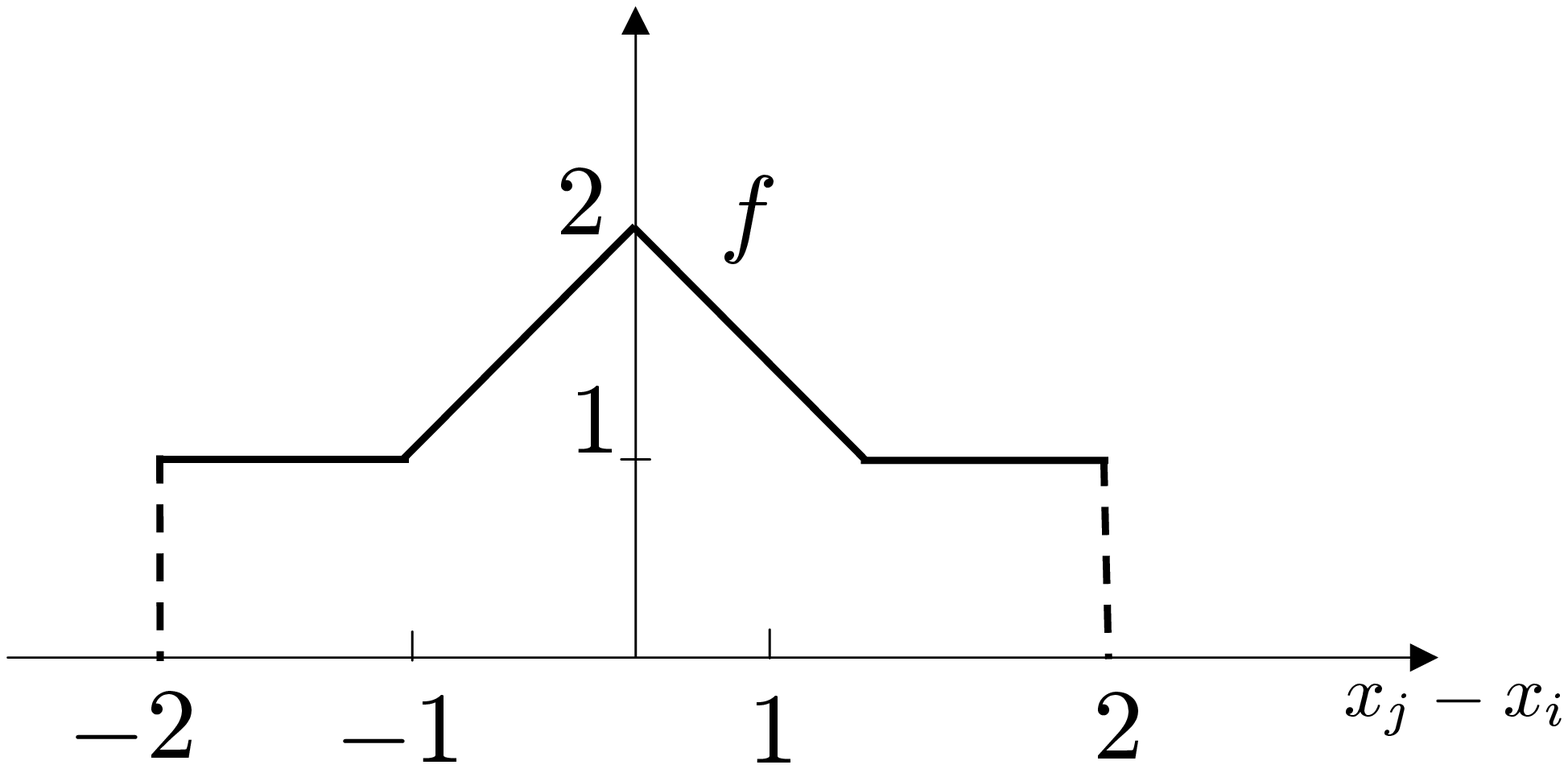}&
\includegraphics[scale=.25]{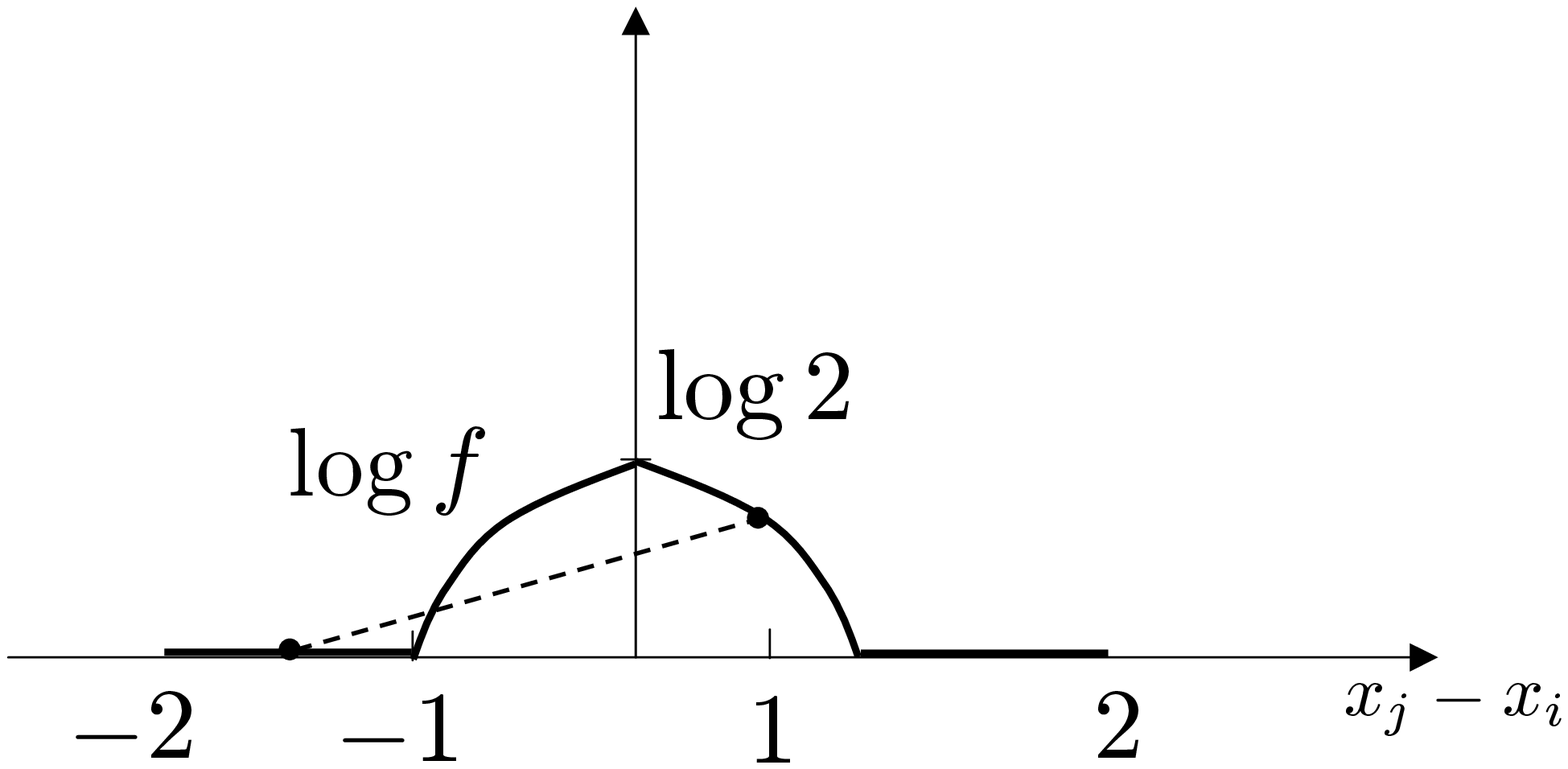}\\(a) &(b)

\end{tabular} \caption{
The function defined by $f(x) = \chi_{[-2,2](x)
}\max\prt{1,1-\abs{x}}$ and represented in (a) is not order
preserving because its logarithm represented in (b) is not
concave, as shown by the dashed line. }\label{fig:ex_plin}
\end{figure}

A consequence of Theorem \ref{thm:logconcave} is that no function
discontinuous on the interior of its domain (and admitting a
positive lower bound on at least one interval) is order
preserving. Such function is indeed never concave. No function
similar to $f_3$ in Figure \ref{fig:various_op_function}, or
containing a gap, is thus order preserving. Similarly, the
function defined by $f(x) = \chi_{-2,2}(x)\max(2-\abs{x},1)$
represented in Figure \ref{fig:ex_plin} is not order preserving
either because its logarithm is not concave. Based on Theorem
\ref{thm:logconcave}, one can actually show that no non-concave
piecewise linear function is order preserving.\\

On the other hand, influence functions such as $\max(1-x^2,0)$ or
$\chi_{[-\pi/2,\pi/2]}(x) \cos(x)$ are concave on their support,
and it follows then from Corollary \ref{cor:concave} that they are
order preserving. For the same reasons, the functions $f_1$ and
$f_2$ in Figure \ref{fig:various_op_function} are also order
preserving. Functions such as $e^{-x^2}$ and even $e^x$ are not
concave, but their respective logarithm $-x^2$ and $x$ are
concave, so that they are also order preserving by Theorem
\ref{thm:logconcave}.

\section{Conclusions and open questions}\label{sec:ccl_openquestions}

We have shown that an influence function is order preserving if it
is log-concave, and that an order preserving function is
log-concave unless it is discontinuous at every point of its
support and admits a positive lower bound on no positive
length-interval. The existence of such order preserving functions
that are not log-concave remains however open. Besides, Krause's
model has also been defined for two or more dimensional spaces, to
which the order preservation property cannot be extended.
Log-concave functions might however have a more generic property
in those spaces, which would imply order preservation for
one-dimensional spaces. Finally, since the order preservation
property is widely used in the mathematical analysis of Krause's
initial model, it would be interesting to see how exactly are
affected the main features of the system when an influence
function is selected.
\\\\

\newpage

\begin{bf} Acknowledgments \end{bf}\\

This research was supported by the Concerted Research Action (ARC)
\quotes{Large Graphs and Networks} of the French Community of
Belgium and by the Belgian Programme on Interuniversity Attraction
Poles initiated by the Belgian Federal Science Policy Office. The
scientific responsibility rests with the author. Julien Hendrickx
holds a F.R.S.-FNRS fellowship (Belgian Fund for Scientific
Research).


\appendix\section{Proof of Proposition \ref{prop:if_not_concave}}\label{sec:appen_proof}

Let $f$ be an order preserving function and call $S_f$ its
support. Suppose that $f$ is not continuous at some $x_0$ in the
interior of $S_f$. We prove that this implies the unboundedness of
$\log f$ on all positive length intervals in $S_f$. In particular,
we show that $\log f$ is unbounded on $[x-\epsilon,x + \epsilon]
\cap S_f$ for any $\epsilon >0$ and $x\in S_f$, and therefore
continuous at no point of $S_f$. This implies that $f$ is also
continuous nowhere on $S_f$, and admits a positive lower bound on
no positive length
interval.\\

\begin{figure}
\centering
\includegraphics[scale = .45]{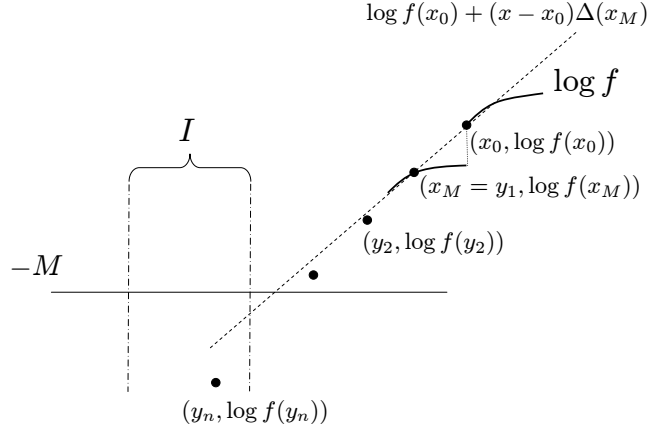}
\caption{Illustration of the construction in the proof of
Proposition \ref{prop:if_not_concave}. All $\log f(y_i)$ must be
below $\log f(x_0) + (y_i - x_0)\Delta(x_M)$.
}\label{fig:repres_constru}
\end{figure}

Let $\Delta(x) = \frac{\log f(x)-\log f(x_0)}{x-x_0}$. The
discontinuity of $\log f$ at $x_0$ implies that $\Delta(x)$ is
unbounded on any open interval containing $x_0$. We suppose that
it takes arbitrary large positive values on any such interval. If
it is not the case, then it necessarily takes arbitrary large
negative values, and a similar argument can be applied. Consider
an arbitrary large $M$ and a interval $I\subset S_f$ of positive
length $\abs{I}$ with $\sup I < x_0$, where by $\sup I$ we denote
the supremum of $I$. The following construction is illustrated in
Figure \ref{fig:repres_constru}. There is a $x_M \in
(x_0-\abs{I},x_0+\abs{I}) $ such that $\abs{\frac{M+\log
f(x_0)}{\sup I - x_0}} < \Delta(x_M)$. Consider now the sequence
of points defined by $y_0=x_0$ and $y_i = y_{i-1} -\abs{x_M-x_0}$.
Since all $y_i$ are smaller than or equal to $x_0$, it follows
from Theorem \ref{thm:orderpreserve} that $\frac{f(y_{i-1})}{
f(y_i)} \geq \frac{ f(x_M)}{f(x_0)}$ holds if $x_M>x_0$ and
$\frac{f(y_{i-1})}{ f(y_i)} \geq \frac{ f(x_0)}{f(x_M)}$ holds if
$x_M<x_0$. In both cases, this implies that $\log f(y_{i-1})- \log
f(y_i) \geq \abs{\log f(x_M)- \log f(x_0)}$ and thus that
\begin{equation*}
\log f(y_i) \leq \log f(x_0) - i \abs{\log f(x_M)- \log f(x_0)} =
\log f(x_0) - (x_0-y_i)\Delta(x_M).
\end{equation*}
Since $\abs{x_M-x_0}< \abs{I}$, there is a $n$ such that $y_n \in
I$. For this $y_n$, there holds $x_0 - y_n \geq \abs{x_0 - \sup
I}\geq \frac{M+\log f(x_0)}{\Delta(x_M)}$. It follows then from
the inequality above that
\begin{equation*}
\log f(y_n)  \leq\log f(x_0) - (x_0-y_n)\Delta(x_M)\leq  \log
f(x_0) - \abs{x_0 - \sup I} \Delta(x_M) <
 -M.
\end{equation*}
Therefore, $\log f$ takes arbitrary large negative values on any
positive length interval $I$ with $\sup I < x_0$.\\

Consider now a $x_1 < x_0$. For any $\delta$, $\log f$ takes
arbitrary large negative values on $[x_1,x_1+\delta]$, and
therefore so does $\Delta_1(x) := \frac{\log f(x)-\log
f(x_1)}{x-x_1}$. It follows then from a similar argument as above
that $\log f$ admits no lower bound on any positive length
interval $I$ with $\inf I > x_1$, and thus that it does not admit
any lower bound on any positive length interval contained in $S_f$
since every such interval contains at least a subinterval $I$
with $\inf I > x_1$ or with $\sup I < x_0$.\\

\begin{thebibliography}{10}

\bibitem{BattistonBonabeauWeisbuch:2003}
S.~Battiston, E.~Bonabeau, and G.~Weisbuch.
\newblock Decision making dynamics in corporate boards.
\newblock {\em Physica A}, 322:567--582, May 2003.

\bibitem{Ben-naimKrapivskyRedner:2003}
E.~Ben-Naim, P.L. Krapivsky, and S.~Redner.
\newblock Bifurcations and patterns in compromise processes.
\newblock {\em Physica D}, 183(3):190--204, 2003.

\bibitem{Ben-naimKrapivskyVasquezRedner:2003}
E.~Ben-Naim, P.L. Krapivsky, F.~Vasquez, and S.~Redner.
\newblock Unity and discord in opinion dynamics.
\newblock {\em Physica A}, 330:99--106, 2003.

\bibitem{BlondelHendrickxTsitsiklis:2007_journal}
V.D. Blondel, J.M. Hendrickx, and J.N. Tsitsiklis.
\newblock A simple multi-agent system with position dependent communication topology.
\newblock {\em preprint}.

\bibitem{BlondelHendrickxTsitsiklis:2007_ECC}
V.D. Blondel, J.M. Hendrickx, and J.N. Tsitsiklis.
\newblock On the 2R conjecture for multi-agent systems.
\newblock In {\em Proceedings of the European control conference 2007 (ECC
  2007)}, pages 2996--3000, Kos, Greece, July 2007.

\bibitem{DeffuantNeauAmblardWeisbuch:2000}
G.~Deffuant, D.~Neau, F.~Amblard, and G.~Weisbuch.
\newblock Mixing beliefs among interacting agents.
\newblock {\em Advances in Complex Systems}, 3:87--98, 2000.

\bibitem{DeGroot:1974}
M.H. DeGroot.
\newblock Reaching a consensus.
\newblock {\em Journal of the American Statistical Association},
  69(345):118--121, 1974.

\bibitem{Dittmer:2001}
J.C. Dittmer.
\newblock Consensus formation under bounded confidence.
\newblock {\em Nonlinear Analysis}, 47:4615--4621, August 2001.

\bibitem{Fortunato:2004}
S.~Fortunato.
\newblock Damage spreading and opinion dynamics on scale-free networks.
\newblock {\em Physica A}, 348:683--690, 2004.

\bibitem{Fortunato:2005_treshold}
S.~Fortunato.
\newblock On the consensus threshold for the opinion dynamics of
  {K}rause{-H}egselmann.
\newblock {\em International Journal of Modern Physics C}, 16(2):259--270,
  2005.

\bibitem{Fortunato:2005}
S.~Fortunato.
\newblock The Sznajd consensus model with continuous opinions.
\newblock {\em International Journal of Modern Physics C}, 16(1):17--24, 2005.

\bibitem{FortunatoLatoraPluchinoRapisarda:2005}
S.~Fortunato, V.~Latora, A.~Pluchino, and A.~Rapisarda.
\newblock Vector opinion dynamics in a bounded confidence consensus model.
\newblock {\em International Journal of Modern Physics C}, 16(10):1535--1551,
  2005.

\bibitem{Galam:1999}
S.~Galam.
\newblock Application of statistical physics to politics.
\newblock {\em Physica A}, 273:132--139, 1999.

\bibitem{HegselmannKrause:2002}
R.~Hegselmann and U.~Krause.
\newblock Opinion dynamics and bounded confidence models, analysis, and
  simulations.
\newblock {\em Journal of Artifical Societies and Social Simulation}, 5(3),
  2002.

\bibitem{Hendrickx:2008phdthesis}
J.M. Hendrickx.
\newblock {\em Graphs and Networks for the Analysis of Autonomous Agent
  Systems}.
\newblock PhD thesis, Université catholique de Louvain,
  http://edoc.bib.ucl.ac.be:81/{ETD}-db/collection/available/BelnUcetd-0211200%
8-142837/unrestricted/Thesis\_Julien\_Hendrickx.pdf, 2008.

\bibitem{KrapivskyRedner:2003}
P.~L. Krapivsky and S.~Redner.
\newblock Dynamics of majority rule in two-state interacting spin systems.
\newblock {\em Physical Review Letters}, 90(23):238701, June 2003.

\bibitem{Krause:1997}
U.~Krause.
\newblock Soziale dynamiken mit vielen interakteuren. eine problemskizze.
\newblock In {\em Modellierung und Simulation von Dynamiken mit vielen
  interagierenden Akteuren}, pages 37--51. 1997.

\bibitem{Krause:2000}
U.~Krause.
\newblock A discrete nonlinear and non-autonomous model of consensus formation.
\newblock {\em Communications in Difference Equations}, pages 227--236, 2000.

\bibitem{Lorenz:2005}
J.~Lorenz.
\newblock A stabilization theorem for continuous opinion dynamics.
\newblock {\em Physica A}, 355(1):217--223, 2005.

\bibitem{Lorenz:2006}
J.~Lorenz.
\newblock Consensus strikes back in the Hegselmann-Krause model of continuous
  opinion dynamics under bounded confidence.
\newblock {\em Journal of Artificial Societies and Social Simulation},
  9(1):http://jasss.soc.surrey.ac.uk/9/1/8.html, 2006.

\bibitem{Lorenz:2007}
J.~Lorenz.
\newblock Continuous opinion dynamics under bounded confidence: A survey.
\newblock {\em International Journal of Modern Physics C}, 18(12):1819--1838, 2007.

\bibitem{Lorenz:2007phdthesis}
J.~Lorenz.
\newblock {\em Repeated Averaging and Bounded Confidence - Modeling, Analysis
  and Simulation of Continuous Opinion Dynamics}.
\newblock PhD thesis, Universität Bremen,
  http://www.informatik.uni-bremen.de/\~{ }jlorenz/Diss\_lorenz.pdf, 2007.

\bibitem{Sznajd:2000}
K.~Sznajd-Seron and J.~Sznajd.
\newblock Opinion evolution in closed community.
\newblock {\em International Journal of Modern Physics C}, 11(6):1157--1165,
  2000.

\bibitem{VicsekCzirolBenjacobCohenSchchet:1995}
T.~Vicsek, A.~Czirok, I.~Ben~Jacob, I.~Cohen, and O.~Schochet.
\newblock Novel type of phase transitions in a system of self-driven particles.
\newblock {\em Physical Review Letters}, 75:1226--1229, 1995.

\bibitem{WeisbuchDeffuantAmblard:2005}
G.~Weisbuch, G.~Deffuant, and F.~Amblard.
\newblock Persuasion dynamics.
\newblock {\em Physica A}, 353:555--575, 2005.

\end{thebibliography}
\end{document}